\documentclass{sig-alternate}

\usepackage{graphicx}
\usepackage{amsmath}

\begin{document}
%
\conferenceinfo{WOODSTOCK}{'97 El Paso, Texas USA}

\title{Resilience of human brain functional coactivation networks under thresholding}
%
%
%
%
%

\numberofauthors{3} 
%
\author{
%
%
\alignauthor
Somwrita Sarkar\\
       \affaddr{Design Lab}\\
       \affaddr{University of Sydney}\\
       \affaddr{NSW 2006, Australia}\\
       \email{\small somwrita.sarkar@sydney.edu.au}
\alignauthor
Sanjay Chawla\\
       \affaddr{School of Information Technologies}\\
       \affaddr{University of Sydney}\\
       \affaddr{NSW 2006, Australia}\\
       \email{\small sanjay.chawla@sydney.edu.au}
\alignauthor
Hanley Weng\\
       \affaddr{Design Lab}\\
       \affaddr{University of Sydney}\\
       \affaddr{NSW 2006, Australia}\\
       \email{\small hanley.weng@sydney.edu.au}
}

\maketitle
\begin{abstract}
Recent studies have demonstrated the existence of community structure and rich club nodes, (i.e., highly interconnected, high degree hub nodes), in human brain functional networks. The cognitive relevance of the detected modules and hubs has also been demonstrated, for both task based and default mode networks, suggesting that the brain self-organizes into patterns of co-activated sets of regions for performing specific tasks or in resting state. In this paper, we report studies on the resilience or robustness of this modular structure: under systematic erosion of connectivity in the network under thresholding, how resilient is the modularity and hub structure? The results show that the network shows show strong resilience properties, with the modularity and hub structure maintaining itself over a large range of connection strengths. Then, at a certain critical threshold that falls very close to 0, the connectivity, the modularity, and hub structure suddenly break down, showing a phase transition like property. Additionally, the spatial and topological organization of erosion of connectivity at all levels was found to be homogenous rather than heterogenous; i.e., no ``structural holes" of any significant sizes were found, and no gradual increases in numbers of components were detected. Any loss of connectivity is homogenously spread out across the network. The results suggest that human task-based functional brain networks are very resilient, where the whole network structure fails only when connectivity is almost fully removed from the network. The findings may help further the understanding of dynamics of and relationships between structural and functional brain networks.  

\end{abstract}




\section{Introduction}
With graph theory, network science, and computational neuroscience joining forces, significant advances have been made in furthering the understanding of human brain network organization~\cite{hagmann2008, bullmore2009, meunier2009, meunier2010, sporns2011, rubinov2013}. Recent studies have demonstrated the existence of small-worldness, community structure and a set of rich club nodes, (i.e., highly interconnected, high degree hub nodes, with large periphery), in human brain structural and functional networks~\cite{crossley2013, meunier2009, meunier2010, hagmann2008}. The cognitive relevance of the detected modules and hubs has also been demonstrated, for both task based and default mode or resting state networks, suggesting that the brain self-organizes into co-activated sets of regions for performing specific tasks or in resting state~\cite{crossley2013}.  

Most of these studies on the functional structure of brain networks, that is, patterns of measured co-activation of brain regions as the brain is in a resting state or performing a specific task, is based on the experimental basis provided by fMRI or functional magnetic resonance imaging studies~\cite{friston2011, sporns2011}. After considerable pre- and post-processing of such data, fMRI based brain networks are produced, on which established graph theoretic algorithms are applied, to deduce information about the topology and spatial organization of brain networks, and the relationship of graph theoretic properties to brain function and dynamics. However, mining and detection of these patterns is dependent on the data, and it is known that in the brain domain, experimentally measured data is subject to innumerable variations on experimental conditions, pre- and post-processing steps, and human subjects, along with combinatorial explanations of these variations~\cite{friston2011}. Therefore, to base the results of analysis on a simple application of algorithms to numerical data can be problematic in terms of interpretation of the data~\cite{friston2011}. Further, while small-worldness, modularity structure, and hub structure has essentially been demonstrated by multiple studies, the resilience of their topological and spatial organization and their relationship to the numerical data provided by the experiments is still an open question. 

In this paper, we report studies on the resilience or robustness of this modular structure: under systematic erosion of connectivity in the network under thresholding, how resilient is the modularity and hub structure? The results show that the functional co-activation network shows strong resilience properties, with the modularity and hub structure maintaining itself over a large range of connection strengths. Then, at a certain critical threshold that falls very close to 0, the connectivity, the modularity, and hub structure suddenly breaks down, showing a phase transition like property. Additionally, the spatial and topological organization of erosion of connectivity at all levels was found to be homogenous rather than heterogenous; no ``structural holes" of any significant sizes appeared, and no gradual increases in numbers of components were detected. Loss of connectivity was homogenously spread out across the network. The results suggest that human task-based functional brain networks are very resilient, where the whole network structure fails only when connectivity is almost fully removed from the network. The findings may help further the understanding of dynamics of functional brain networks, such as their criticality and stability, ad the relationships between structure and function. 

\subsection{Paper summary}
We first present a static modularity analysis and rich club or hub analysis as presented in previous work~\cite{crossley2013}. Our results match with the previously reported results in the literature~\cite{crossley2013}. Then, in Section~\ref{Sec2} we present our thresholding method, and its application on the data to produce graphs at various levels of decreasing connectivity. In Section~\ref{Sec3} we analyze the dynamics of how the community structure organization and rich club structure changes dynamically as the connectivity in the graph is systematically eroded. We show the existence of a sudden phase transition like behavior, where the community structure organization suddenly breaks down, and we show that this is in the region that represents the bulk of connectivity in the network. We also analyze how the rich club structure changes. Finally, in Section~\ref{Sec4} we summarize our findings, discuss our results, and present ideas for future work.

\section{Data and thresholding}
\label{Sec2}

From~\cite{crossley2013}, we have obtained data on functional coactivation brain networks. In the original paper, the authors have collected 1641 instances of task-related fMRI or PET studies published between 1985 and 2010, available in the BrainMap database. The human brain has been then parcellated into 638 regions covering the whole brain, and for every pair of these regions, a Jaccard Index similarity was defined, that was computed using the 1641 instances of task related data mentioned above. That is, for every individual task, patterns of co-activation between pairs of brain regions were available, and the Jaccard Index was then used as meta-analysis, to provide a measure of "co-activation" association between every two pairs of 638 brain regions over the entire set of tasks. This provides the functional co-activation matrix, where a high value of co-activation between two brain regions indicates the high measure of similarity between the patterns of activations of two regions reported across a range of task-based experiments. For more information, please see the original source~\cite{crossley2013}. 

Using the functional co-activation matrix, a network $G$ can be defined with $V$ vertices, and $E$ edges, where two vertices $v_{i}$ and $v_{j}$ are connected by an edge with a weight $w_{ij}$ equal to the functional co-activation measure described above. If two nodes do not have an edge between them, this means that the corresponding brain regions have no co-activation. From the data available, the numerical range is particularly small, varying from 0 (no co-activation) to 0.3355 (highest co-activation). The adjacency matrix is represented by $A$, where $A_{ij} = w_{ij}$ if the nodes are connected, and $A_{ij}=0$ otherwise. 

The original source~\cite{crossley2013} reports that this graph has a strong modular structure with four principal modules detected (occipital, central, frontoparietal, and default mode), that not only correspond to the known large scale structural organization of the brain, but also correspond to the known specific behavioral, and cognitive functions supported by these regions in the brain, thereby showing the relationship between structural and functional structure. We reproduced these results, by applying the same algorithm to the data as the authors, and confirmed the existence of these modules. Additionally, a rich-club structure was also reported in the study, showing the existence of a set of highly connected, highly interrelated, high degree hub nodes, Therefore, it was interesting to ask the following question: how resilient is this network to systematic erosion of connectivity? The answer to this question would not only help to deepen our understanding of the topological and spatial organization of the modularity and hub structure of the functional network, it would also help us to understand the patterns of co-activation as they are spread across the brain, both topologically and spatially. More specifically, we ask the classic \textit{integration} versus \textit{segregation} question: under systematic erosion of connectivity, does the co-activation network fracture into separate components gradually with modularity structure rapidly changing, or does the modularity structure hold as network connectivity erodes, pointing to a more globally integrated structure of co-activation. 

We did the following. Looking at the data range and distribution, we identified a set of relevant numerical thresholds $t$. Then, at every step, a new network was produced from the original network, with all edges above the value of $t$ being removed from the network. Repeated application of this process produced a set of networks with the same set of nodes, but decreasing in connectivity. This process was performed till there were no edges left in the network. These new sparse networks were then examined for connectivity, modularity and hub (degree centrality) structure. For modularity, we used the Louvain algorithm that the authors of the original paper used. The Louvain algorithm is a well known fast algorithm for Newman's modularity maximization~\cite{blondel2008}.  

\section{Results}
\label{Sec3}

\subsection{Modularity and hubs analysis}

\begin{figure*}
\centering
\includegraphics[width=\textwidth]{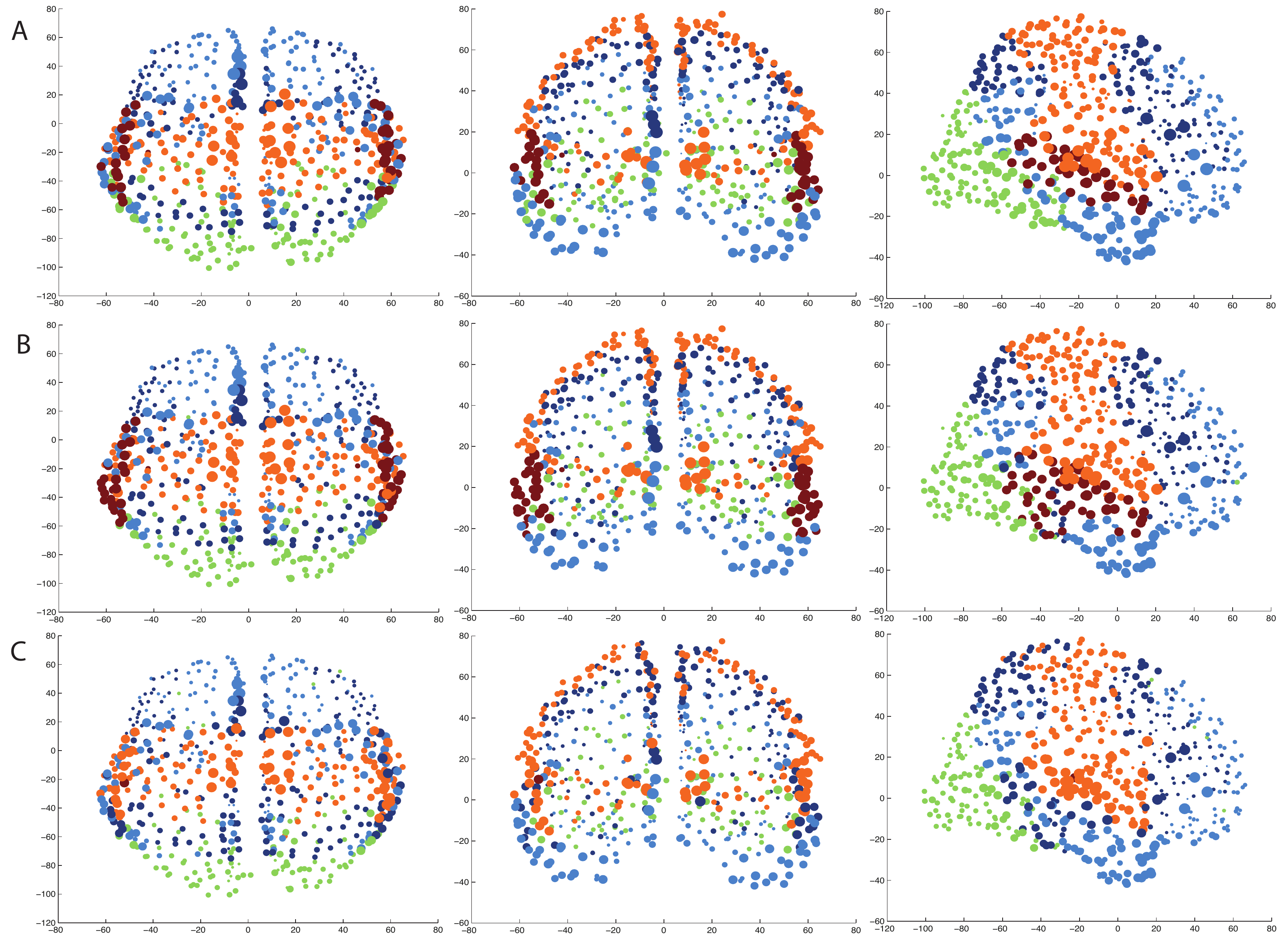}
\caption{A. Coactivation network modules found by Newman's modularity maximization using the Louvain algorithm plotted in anatomical space for human functional coactivation networks. Modularity value Q = 0.474, 5 clusters obtained by running the Louvain algorithm, each plotted with a distinct color; results are similar to \cite{crossley2013} from where we have obtained the data. Modules: Green: Occipital, Orange and dark red: Central, Dark Blue: Frontoparietal, Light blue: Default mode. B and C. Original network has been thresholded at 0.2 and 0.1 respectively; i.e., i.e., all connections higher than the threshold have been removed from the network and modularity and degrees re-computed. The edges have not been plotted for clarity of representation.}
\label{Fig1}
\end{figure*}

Figure~\ref{Fig1}A shows the results of the modularity analysis for the original functional coactivation network. We detected 5 large clusters, corresponding exactly with the 4 large clusters reported in the original source~\cite{crossley2013}. The 4 detected modules were largely identified to be the occipital, central, fronto-parietal, and default mode. In our case, two clusters were detected for the fronto-pareital, making the total number of modules 5.  

\begin{figure}
\centering
\includegraphics[width=0.4\textwidth]{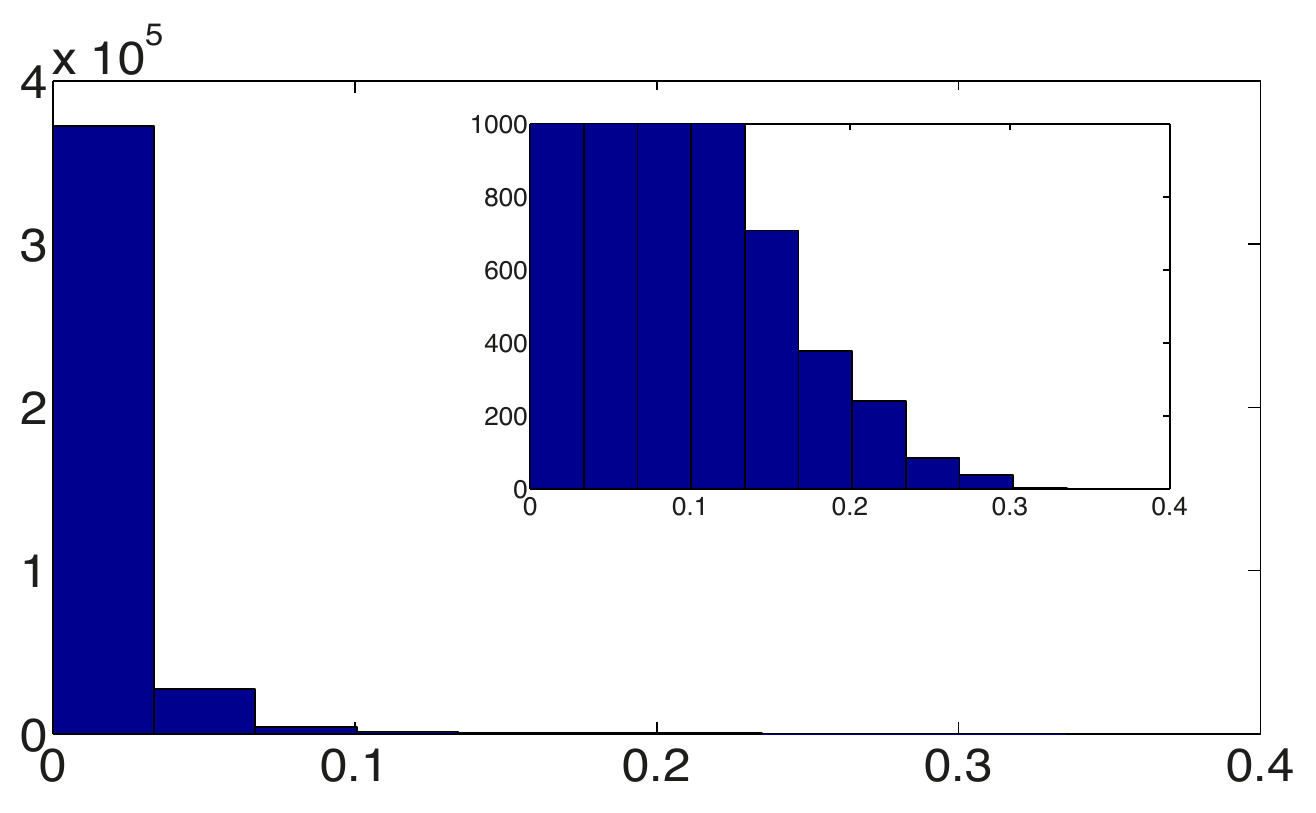}
\caption{The distribution of edge weights in the human functional co-activation network}
\label{Fig2}
\end{figure}

\begin{figure*}
\centering
\includegraphics[width=\textwidth]{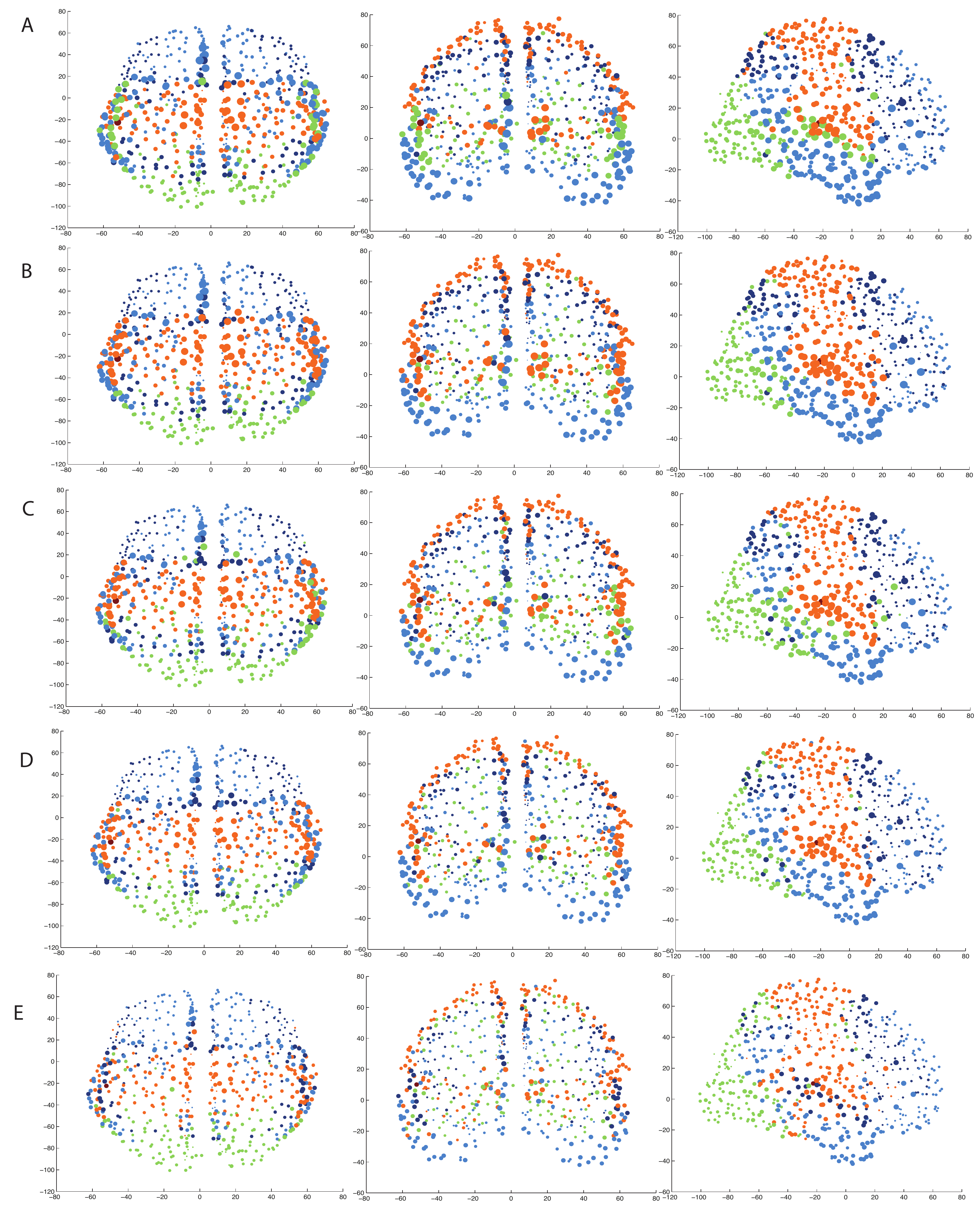}
\caption{Modularity and hubs in human brain functional coactivation network under systematic thresholding of connectivity. Colors represent modules. Sizes of circles represent degrees of nodes. Edges have not been shown for visual clarity, since these are extremely dense networks. A, B, C, D, and E show the modularity and hub locations for networks thresholded at 0.09, 0.08, 0.07, 0.06, and 0.05 strengths of connectivity respectively; i.e., all connections higher than the threshold have been removed from the network and modularity and degrees computed. Note the gradual homogenous decay upto the threshold point (0.04-0.03) beyond which the whole network breaks down suddenly, and just above the threshold, (till 0.05), the network maintains its connectivity and community structure, even though the network becomes sparser.}
\label{Fig3}
\end{figure*}

We then performed an analysis of the distribution of edge strengths in the network~\ref{Fig2}. The data range is small, and varies from 0 to a maximum of 0.3355, but most of the connectivity is shrunk into an even much smaller range, between 0 and 0.1. Therefore, we chose thresholds sensitive to this distribution of data to be 0.2, 0.1, 0.09, 0.08, 0.07, 0.06, 0.05, 0.04, 0.03, 0.02, and 0.01. Figure~\ref{Fig1}B and C, and Figure~\ref{Fig3} show the results of modularity and hub analysis for all these thresholds, showing the gradual decay of the network connectivity, but the persistence of the rich club and the modularity structure. 

\begin{figure*}
\centering
\includegraphics[width=\textwidth]{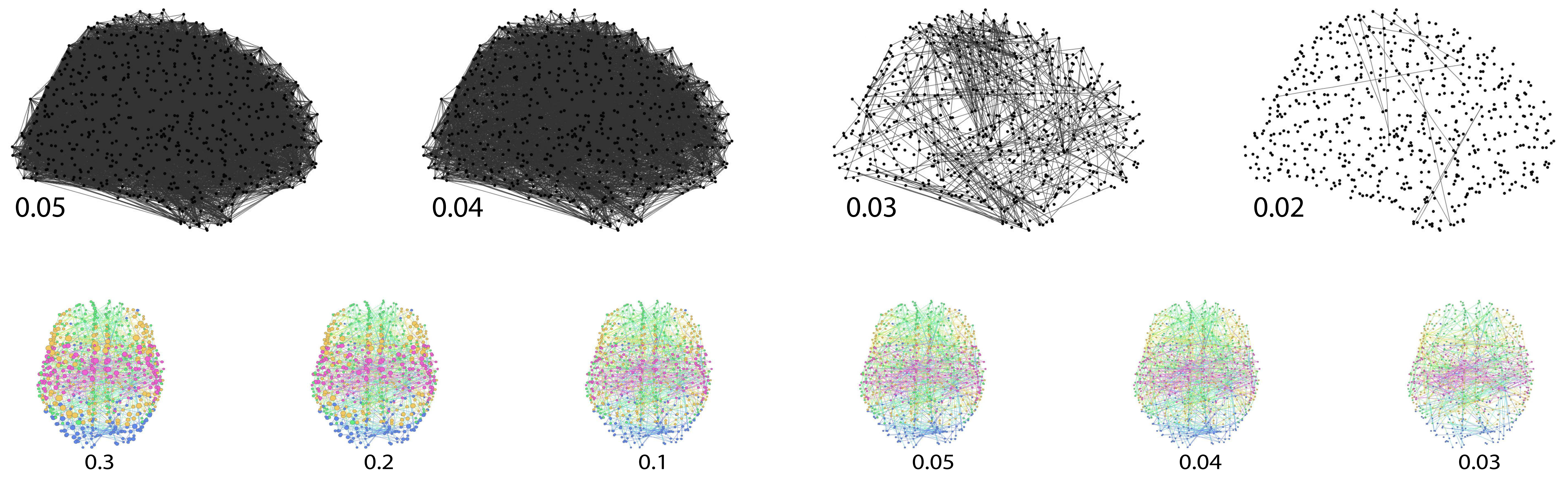}
\caption{Visualization of the phase transition like behavior: at 0.05 the network is dense, at 0.04 two components are formed, but the network is still dense, at 0.03 connectivity is suddenly lost, and by 0.02 almost no edges are left in the network. The frame below shows the visualization of degrees with the sizes of circles representing degrees, by 0.05, 0.04 or 0.03, the original rich club nodes go down in degree, but till 0.05 the network is still fully connected. The minimum spanning tree of the netwotk has been used for ease of visualization in the frame below.}
\label{Fig3a}
\end{figure*}

At 0.03, the connectivity and modularity and hub structure suddenly breaks down, showing a phase transition like behavior. That is, till the thresholded range of 0.05, the modularity analysis and rich club structure continue to show the same results; i.e., 4 main modules, and persistent nodes with high degrees as hubs. However, suddenly, at 0.04-0.03, the modularity structure suddenly breaks with more than 300 modules emerging at 0.03. To test this finding more deeply, we tracked the Laplacian eigenvalues and their relationship to the Cheeger constant, tracking the appearance of connected components and the ease of partitioning a network into subgraphs (see below). 

\begin{figure}
\centering
\includegraphics[width=0.4\textwidth]{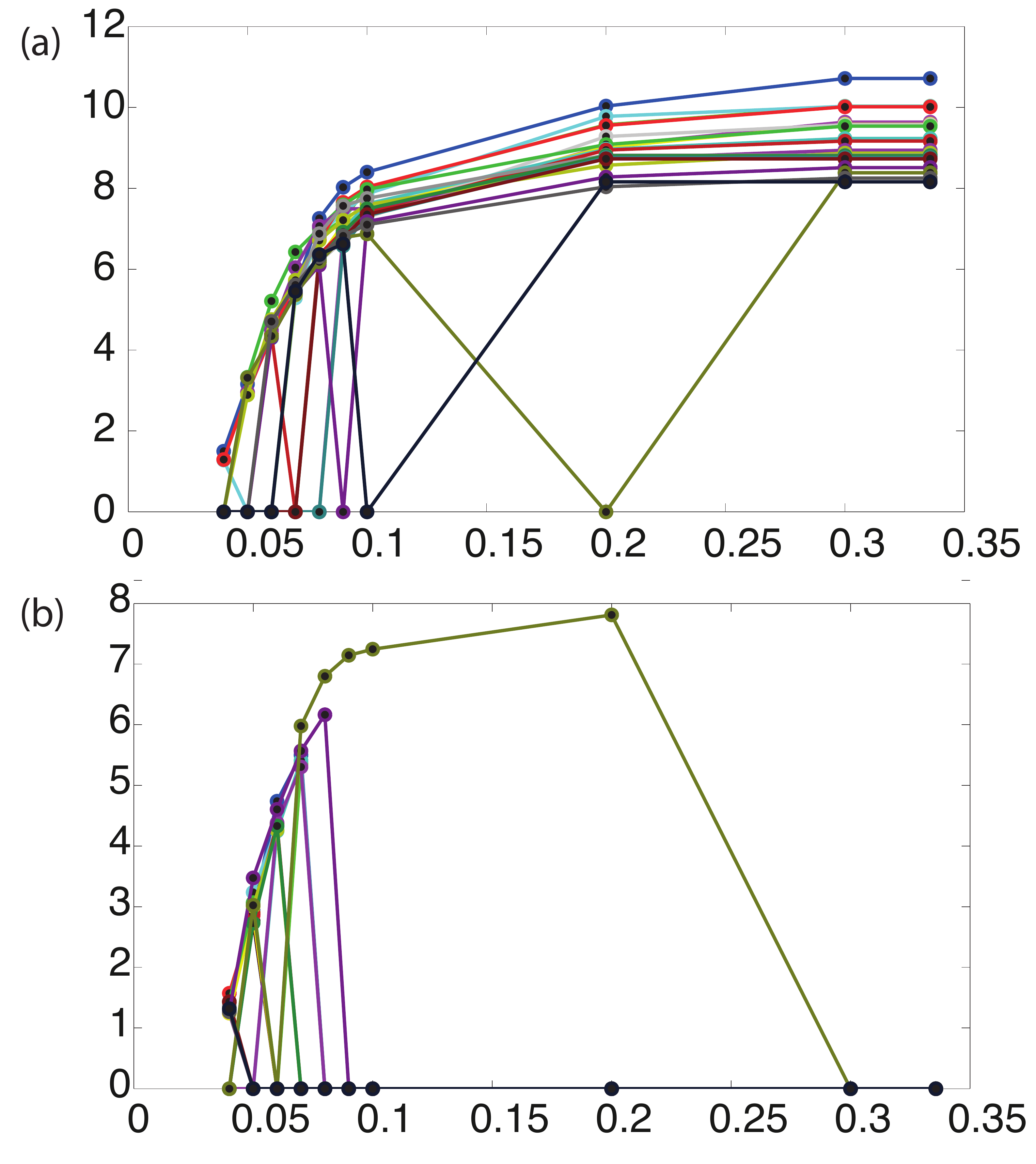}
\caption{Analysis of the rich club nodes for original and thresholded networks. The x-axis shows the threshold values, and the y-axis tracks the degrees of the nodes that appear in the rich club (highest 21 nodes by degree) at each level of thresholding. Lines are for visual tracing convenience only. (a) the 21 top nodes as reported in~\cite{crossley2013}. (b) other nodes that appear in the top 21, especially in the thresholding zone 0.1 to 0.05.}
\label{Fig3b}
\end{figure}

Further, the authors in~\cite{crossley2013} also reported the existence of 21 principal nodes that form the rich club of hubs. These are characterized by a high degree of interconnectedness between themselves, as well as the rest of the network, and have the highest degrees. We tracked 50 nodes with the highest degree centrality in each of the original and thresholded network versions, and tracked the appearance and disappearance of nodes from this ``rich club" set at every thresholding step. Figure~\ref{Fig3b}a shows the results of the analysis: the top 21 nodes with the highest degree show high resilience to erosion over a large range of thresholding, and at a threshold value of 0.1-0.04 again the same phase transition like behavior is seen. Their degrees suddenly drop drastically in this range. The lines have been plotted for visual tracking convenience, and the points where degree is shown to be zero do not signify that the degree has dropped to zero, rather it signifies that that particular node has moved out of the ``rich club" of the 21 highest degree nodes at that level of thresholding. Interestingly, some of the nodes go out of the rich club and then appear again. Since we tracked the top 50 nodes for their degrees at every level of thresholding, Fig.~\ref{Fig3b}b shows a few nodes that  appear in the rich club in the zone of thresholding where the nodes of the original rich club move out of the rich club: 0.1 to 0.05.

\subsection{Zeros of Laplacian}
\begin{figure}
\centering
\includegraphics[width=0.4\textwidth]{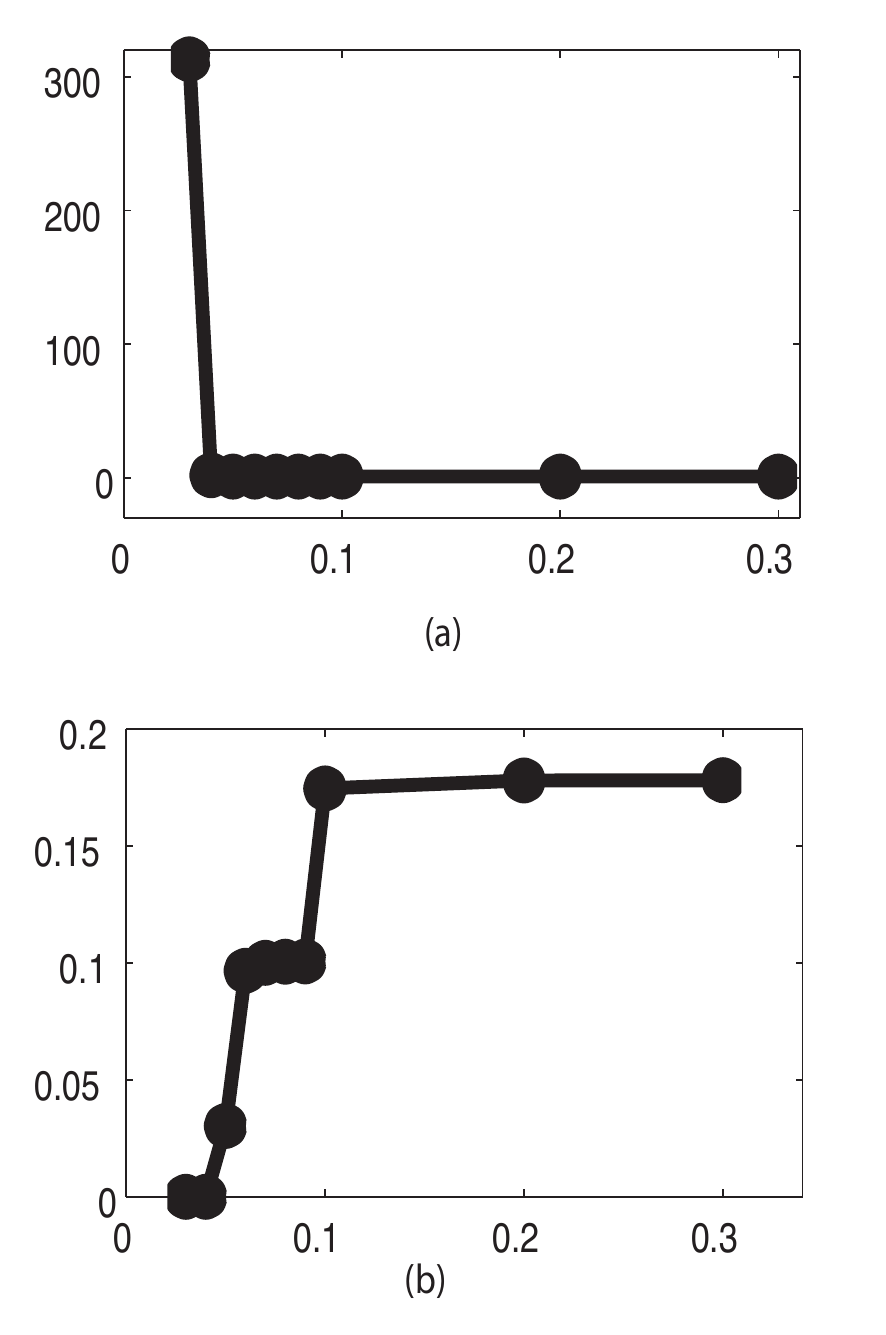}
\caption{Tracing the first and second smallest eigenvalues of the Laplacian matrices of thresholded networks to chart number of connected components of the network. The x-axis shows the thresholds and the y-axis shows (a) the number of zero eigenvalues for each thresholded matrix, and (b) the actual value of the second smallest eigenvalue.}
\label{Fig4}
\end{figure}

The Laplacian is defined as $L = D - A$, where $D$ is a diagonal matrix with $d_{ii}$ equal to the degree of the node $i$. A well-known theorem states that the first eigenvalue of the Laplacian is always a zero, and the number of zeros in the eigenspectrum signifies the number of disconnected, connected components in the graph. Figure~\ref{Fig4} shows the results of this analysis, confirming the observations above. Only the first eigenvalue is a zero till the threshold of 0.05, at 0.04 there are two zero eigenvalues, signifying 2 components, and thereafter, at 0.03, there are 313 zeros, Fig.~\ref{Fig4}a. Further, tracking the second smallest eigenvalue next to zero,Fig.~\ref{Fig4}b, shows that it is positive but falling till 0.05, and then falls to zero at 0.04. 

\subsection{Cheeger constant}
Given any two vertex-disjoint subgraphs $V_{1}$ and $V_{2}$, we let $E(V_{1}, V_{2})$ represent the set of edges that run between the two vertex groups. Given a graph $G$ and a subgraph $V$ of $G$, we represent by $V'$ the complement of $V$, that is the set of vertices in $G$ but not in $V$. Then, we define \begin{equation} h_{G}(V) = \frac{|E(V,V')|}{min \{\sum_{v\in V}d_v, \sum_{v'\in V'}d_v'\}}, \end{equation} and the \textit{Cheeger constant} is defined as \begin{equation}  h_{G} = min\{h_{G}(V)\}. \end{equation} The Cheeger constant measures the ease with which a graph can be cut into two parts. That is, if it is small, then the number of edges running between a subgraph and its complement is much lower than the degrees of the nodes in the subgraph, and therefore, it will be easy to partition the graph into parts. Conversely, if it is high, then it will be harder to partition the graph into parts, since there is a high number of edges running between subgraphs. Thus, for the brain network, we expect to see high values of the constant $h_{G}(V)$, since it appears that it is difficult to produce components by eroding the connectivity, for a large range of the connectivity strengths or edge weights. 

While it is infeasible to compute the Cheeger constant, we make two observations on approximations. First, it is reasonable to expect that out of all possible subgraphs of a graph $G$, detected modules will have the least values for $h_{G}(V)$, because by definition, modularity finding through modularity maximization finds communities that minimizes the number of edges running between these communities. In other words, by definition, modularity finding means looking for solutions with highest possible intramodule connectivity and lowest possible intermodule connectivity. Thus, computing $h_{G}(V)$ by setting $V$ to be the subgraphs defined by the detected modules will ensure that we are looking for a set with the least possible Cheeger constant values, out of all possible subgraphs in $G$. Second, the second eigenvalue of the Laplacian is computable, and the following relation provides upper and lower bounds on the second eigenvalue of the Laplacian using the Cheeger constant: \begin{equation} \label{eq3} 2h_{G} >= \lambda_{2} >= \frac{h_{G}^{2}}{2}. \end{equation} The second eigenvalue of the Laplacian is also used as a measure of connectivity, via its Fiedler eigenvector, frequently used in spectral graph partitioning. Thus, putting together the above two ideas, we computed the constant $h_{G}(V)$ for each of the subgraphs defined by the communities for the original and thresholded versions of the network and then chose the minimum of these values. Table 1 shows the results. It is worth mentioning that the inequality in Eq~\ref{eq3} is satisfied even with this rough computation. The results in Table 1 show that the constants are large, suggesting that there is a high number of edges running between each subgraph and its complement, implying that the network is not partitioned easily into subgraphs, and is resilient at all levels of thresholding upto the phase transition like point where everything breaks up. For all computations, we have used the weighted information, and not binarized versions of the networks. 

\begin{table*}[ht]
\caption{Constant $h_{G}(V)$ computed for communities in thresholded networks} 
\centering
\begin{tabular}{ | p{1.5cm} | p{1.5cm} | p{1.5cm} | p{1.5cm} | p{1.5cm} | p{1.5cm} || p{1.5cm} | p{1.5cm} | p{1.5cm} |}
\hline
Threshold & Fronto-pareital & Default Mode & Occipital & Central & Central & $2h_{G}$ & $\lambda_{2}$ & $h_{G}^{2}/2$ \\ \hline
Original & 0.6159 & 0.1417 & 0.2965 & 0.2679 & 0.2241 & 0.2835 & 0.1780 & 0.0100 \\ \hline
0.2 & 0.6147 & 0.1568 & 0.3659 & 0.2639 & 0.2033 & 0.3137 & 0.1779 & 0.0123  \\ \hline
0.1 & 0.5345 & 0.1626 & 0.4476& 0.2439 & - & 0.3251 & 0.1745 & 0.0132 \\ \hline
0.09 & 0.8149 & 0.1307 & 0.4005 & 0.3572 & - & 0.2613 & 0.1006 & 0.0085 \\ \hline
0.08 & 0.7801 & 0.1323 & 0.4496 & 0.2295 & - & 0.2646 & 0.1005 & 0.0088 \\ \hline
0.07 & 0.7176 & 0.2457 & 0.3734 & 0.2486 & - & 0.4914 & 0.1000 & 0.0302 \\ \hline
0.06 & 0.6978 & 0.1506 & 0.4995 & 0.2671 & - & 0.3012 & 0.0996 & 0. 0113 \\ \hline
0.05 & 0.7124 & 0.2366 & 0.4602 & 0.3682 & - & 0.4732 & 0.0303 & 0.0280 \\ \hline 
\hline
\end{tabular}
\end{table*}

\section{Conclusions and discussion}
\label{Sec4}
In this paper, we reported studies on the resilience or robustness of the modularity and hub structure of human brain functional co-activation networks. We investigated the question: under systematic erosion of connectivity in the network under thresholding, how resilient is the modularity and hub structure? The results show that modularity and hub structure show strong resilience properties, with the modularity and hub structure maintaining itself over a large range of connection strengths. Then, at a certain critical threshold that falls very close to 0, the connectivity, the modularity, and hub structure suddenly break down, showing a phase transition like property within a very small range of connectivity. Additionally, the spatial and topological organization of erosion of connectivity at all levels was found to be homogenous rather than heterogenous. This implies that no ``structural holes" of any significant sizes were detected and no gradual increases in numbers of components were detected. Our hypothesis thus says that when such holes emerge under loss of connectivity, they are homogenously spread out across the network. This is very different from other networks, such as road and air traffic and transportation networks, for example, where the loss of functional or flow connectivity is much more heterogenous~\cite{batty2013}. The results suggest that human task-based functional brain networks are very resilient, where the whole network structure fails only when connectivity is almost fully removed from the network. 

A number of directions for future work are possible. First, other types of systematic erosion of connectivity could be explored. For example, the same question could be asked by changing the thresholding method to removing the highest degree nodes one by one, and observing how the network structure erodes. Second, the existence of the phase transition like behavior needs to be explored more completely. While there is a lot of reported literature on phase transitions in random and small world networks~\cite{watts1998}, this literature is not directly related to modularity, even though it is now established that all modular networks are small world networks~\cite{pan2009} (although the converse may not be true). Therefore, it will be interesting perform such studies over other data sets, but also model such behavior for modular networks, and relate it to the eigenstructure of the networks. Finally, it will be useful to extend these aspects into exploring the possible implications for nderstanding the dynamics of and relationships between structural and functional brain networks~\cite{robinson2009, robinson2012, sporns2011}. In particular, it is known that higher level functional processes such as information processing in the brain are related to the anatomical structure, but recent studies have shown that this relationship is not only dependent on direct structural connectivity, but also indirect longer paths through the network~\cite{robinson2012}. If indirect paths are important, then the study of how resilient they are at different hierarchical levels will be an interesting question to study. Further, the findings and future studies may help further the understanding of dynamics of functional brain networks, such as their criticality and stability properties at various levels of connectivity and thresholding~\cite{robinson2009}.

\bibliographystyle{abbrv}
\bibliography{sigproc}  

\begin{thebibliography}{10}

\bibitem{batty2013}
M.~Batty.
\newblock {\em The New Science of Cities}.
\newblock MIT Press, 2013.

\bibitem{blondel2008}
V.~Blondel, J.~Guillaume, R.~Lambiotte, and E.~Lefebvre.
\newblock Fast unfolding of communities in large networks.

\bibitem{bullmore2009}
E.~T. Bullmore and O.~Sporns.
\newblock Complex brain networks: graph theoretical analysis of structural and
  functional systems.
\newblock {\em Nat. Rev. Neurosci.}, 10:186--198, 2009.

\bibitem{crossley2013}
N.~Crossley, A.~Mechelli, P.~Vertes, T.~Winton-Brown, A.~Patel, C.~Ginestet,
  P.~McGuire, and E.~Bullmore.
\newblock Cognitive relevance of the community structure of the human brain
  functional co-activation network.
\newblock {\em PNAS}, 110(38):11583--11588, 2013.

\bibitem{friston2011}
K.~Friston.
\newblock Functional and effective connectivity: A review.
\newblock {\em Brain Connectivity}, 1(1):13--36, 2011.

\bibitem{hagmann2008}
P.~Hagmann, L.~Cammoun, X.~Gigandet, R.~Meuli, C.~J. Honey, and V.~J. Weeden.
\newblock Mapping the structural core of the human cerebral cortex.
\newblock {\em PLOS Biol.}, 6(7):e159, 2008.

\bibitem{meunier2010}
D.~Meunier, R.~Lamboitte, and E.~T. Bullmore.
\newblock Modular and hierarchically modular organization of brain networks.
\newblock {\em Front. Neurosci.}, 4:Article 200, 2010.

\bibitem{meunier2009}
D.~Meunier, R.~Lamboitte, A.~Fornito, K.~Ersche, and E.~T. Bullmore.
\newblock Hierarchical modularity in human brain functional networks.
\newblock {\em Front. Neuroinform.}, 3:Article 37, 2009.

\bibitem{pan2009}
R.~K. Pan and S.~Sinha.
\newblock Modularity produces small world networks with dynamical time-scale
  separation.
\newblock {\em Europhys. Lett.}, 85:68006, 2009.

\bibitem{robinson2012}
P.~A. Robinson.
\newblock Interrelating anatomical, effective, and functional brain
  connectivity, using propagators and neural field theory.
\newblock {\em Phys. Rev. E}, 85:011912, 2012.

\bibitem{robinson2009}
P.~A. Robinson, J.~A. Henderson, E.~Matar, P.~Riley, and R.~T. Gray.
\newblock Dynamical reconnection and stability constraints on cortical network
  architecture.
\newblock {\em Phys. Rev. Lett.}, 103:108104, 2009.

\bibitem{rubinov2013}
M.~Rubinov and E.~Bullmore.
\newblock Schizophrenia and abnormal network hubs.
\newblock {\em Dialogues in Clinical Neuroscience}, 15:339--349, 2013.

\bibitem{sporns2011}
O.~Sporns.
\newblock {\em Networks of the brain}.
\newblock MIT Press, 2011.

\bibitem{watts1998}
D.~J. Watts and S.~Strogatz.
\newblock Collective dynamics of small world networks.

\end{thebibliography}
\end{document}